\documentclass[prl,onecolumn,preprint,showpacs,groupedaddress,superscriptaddress,nofootinbib,floatfix,preprintnumbers,longbibliography]{revtex4-2}

\usepackage{physics}
\usepackage{xcolor}
\usepackage{graphicx}
\usepackage{subfigure} 
\usepackage{dcolumn}
\usepackage{bm}

\begin{document}

\title{
Twisting attosecond pulse trains by amplitude-polarization IR pulses}

\author{Enrique G. Neyra}
\email{enriquen@ciop.unlp.edu.ar}
   \affiliation{Centro de Investigaciones \'Opticas (CICBA-CONICET-UNLP), Cno.~Parque Centenario y 506, P.O. Box 3, 1897 Gonnet, Argentina}

\author{Fabi\'an Videla}
\affiliation{Centro de Investigaciones \'Opticas (CICBA-CONICET-UNLP), Cno.~Parque Centenario y 506, P.O. Box 3, 1897 Gonnet, Argentina}
\affiliation{Departamento de Ciencias B\'asicas, Facultad de Ingenier\'ia UNLP, 1 y 47 La Plata,Argentina}

\author{Demian A. Biasetti}
\affiliation{Centro de Investigaciones \'Opticas (CICBA-CONICET-UNLP), Cno.~Parque Centenario y 506, P.O. Box 3, 1897 Gonnet, Argentina}

\author{Marcelo F. Ciappina}
\email{marcelo.ciappina@gtiit.edu.cn}
\affiliation{Department of Physics, Guangdong Technion - Israel Institute of Technology, 241 Daxue Road, Shantou, Guangdong, China, 515063}
\affiliation{Technion -- Israel Institute of Technology, Haifa, 32000, Israel}
\affiliation{Guangdong Provincial Key Laboratory of Materials and Technologies for Energy Conversion, Guangdong Technion - Israel Institute of Technology, 241 Daxue Road, Shantou, Guangdong, China, 515063}
\date{\today}
             
\author{Lorena Reb\'on}
\email{rebon@fisica.unlp.edu.ar}
\affiliation{ 
Departamento de F\'isica, FCE, Universidad Nacional de La Plata, C.C. 67, 1900 La Plata, Argentina
}%
\affiliation{Instituto de F\'isica de La Plata, UNLP - CONICET, Argentina}

\begin{abstract}

Natively, atomic and molecular processes develop in a sub-femtosecond time scale. In order to, for instance, track and capture the electron motion in that scale we need suitable `probes'. Attosecond pulses configure the most appropriate tools for such a purpose. These ultrashort bursts of light are generated when a strong laser field interacts with matter and high-order harmonics of the driving source are produced. In this work, we propose a way to twist attosecond pulse trains. In our scheme, each of the attosecond pulses in the train has a well-defined linear polarization, but with a different polarization angle between them. To achieve this goal, we consider an infrared pulse with a particular polarization state, called amplitude polarization. This kind of pulse was experimentally synthesized in previous works. Our twisted attosecond pulse train is then obtained by nonlinear driving an atomic system with that laser source, through the high-order harmonics generation phenomenon. We reach a high degree of control in the polarization of the ultrashort coherent XUV-generated radiation. Through quantum mechanical simulations, supplemented with signal processing tools, we are able to dissect the underlying physics of the generation process. We are confident these polarized-sculpted XUV sources will play an instrumental role in future pump-probe-based experiments.


\end{abstract}

\maketitle

Classical electrodynamics and quantum mechanics are the fundamental building blocks for the description of many natural phenomena. By measuring the wavelength and speed of light, electromagnetism provides us with tools to extract the velocity of the light field oscillations. Likewise, quantum mechanics links the rapidity of electronic motion with the energy distribution of the populated quantum states. By adequately tuning the light sources, these states can be accessed by photon absorption and emission. The native scale of both the light oscillations and electron dynamics falls within the attosecond range. These elementary processes comprise the constitutional steps of any change in the physical, chemical, and biological properties of materials and soft matter. The capability of capturing and manipulating them in real-time is therefore relevant for the development of novel materials and technologies, as well as the understanding of fundamental atomic and molecular phenomena initiated by light fields~\cite{krauszIvanov,krauszPhysScr2016}.

Since the first measurement of an attosecond pulse train (APT)~\cite{paul2001observation},
attosecond science has grown enormously, from the obtaining of an isolate attosecond pulse (IAP)~\cite{sansone2006isolated} to the generation of optical attosecond pulses~\cite{hassan2016optical}, producing a great deal of applications based on these sources~\cite{krausz2014attosecond,li2020attosecond,midorikawa2022progress}. As is well known, APTs are obtained from the high-order harmonics generation (HHG) phenomenon. Here, a high-intensity infrared (IR) laser pulse interacts with a gaseous system producing, at every half laser cycle, a burst of extreme ultraviolet (XUV) radiation, i.e.~a train of ultrashort XUV pulses is generated. The underlying physics behind the production of APT is rooted in the so-called three-step model, namely (i) tunneling ionization, (ii) classical motion, and (iii) recombination. The energy gained by the electron in its journey through the laser continuum is converted into a high-energy photon, upon recombination with the parent ion. The strong field approximation (SFA) can be used to model the HHG process, and, consequently, the generation of APT~\cite{aminiROPP}. Full-blown quantum mechanical models, based on the numerical solution of the time-dependent Schr\"odinger equation (TDSE) can be utilized as well, although the transparent link between the electron dynamics and their associated fundamental processes is lost.

The ability to obtain IAPs is key to exploring the electron dynamics in its natural time scale~\cite{sansone2006isolated}. Although experimentally it is significantly more demanding to generate an IAP than an APT, different techniques, namely amplitude gating, lighthouse, ionization gating, and distinct polarization gating setups~\cite{timmers2016polarization,calegari2016advances}, can be used to isolate a single attosecond pulse.
On the other hand, manipulating the spatio-temporal properties or polarization state of laser pulses, allows other degrees of freedom to be added to the laser-matter interaction, beyond the frequency and amplitude of the radiation. To achieve this goal, there exists a great variety of pulse-shaping techniques~\cite{misawa2016applications,weiner2011ultrafast,shen2022roadmap} which require the use of optical elements and devices such as wave-plates, spatial light modulators, acousto-optic modulators, interferometer systems, among others, that can work in the visible and IR region of the electromagnetic spectrum. However, to have control of the XUV coherent radiation as in the visible-IR region is highly challenging because, in general, there are no simple physical devices designed for that wavelength range. Therefore, the control of the different properties of the coherent XUV radiation generated via HHG should come from the manipulation of the IR driving pulse.

Several schemes and sources have been used so far, that allow a high degree of control in the produced XUV radiation, e.g.~attosecond light vortices~\cite{hernandez2013attosecond,geneaux2016synthesis}, structured light beams carrying optical angular momentum (OAM)~\cite{rego2019generation}, controlled OAM sources~\cite{minneker2021toru,fang2021controlling}, chirality in nonlinear optics~\cite{neufeld2018optical} and others~\cite{huang2018polarization,watzel2020multipolar}. In particular, by using bichromatic and counter-rotating circularly polarized pulses,
it is possible to generate high-order harmonics 
with circular polarization, in a gaseous medium~\cite{milovsevic2000attosecond,milovsevic2000generation,fleischer2014spin,kfir2015generation}. In the temporal domain, such harmonics
 originate an APT where the  polarization state of each attosecond pulse depends on the Lissajous curves that describe the bichromatic driving field, where these curves are shaped like a flower whose petals are distributed equidistantly on a circumference. 
  Moreover, the temporal structure of these attosecond pulses has a period that depends on the frequency ratio in the bichromatic field~\cite{rego2020trains,jimenez2018attosecond,fleischer2014spin,dorney2017helicity,jimenez2018control,milovsevic2018control}. Such ratio also gives the number of petals in the Lissajous curves and the angle between them. This possibility of manipulating the polarization state of each attosecond pulse in the APT allows for a more sophisticated level of development and understanding of ultrafast magnetism~\cite{bandrauk2017circularly} in the XUV region, both in molecules~\cite{yuan2013attosecond} and in solid systems~\cite{fan2015bright}. In addition, it enables the exploration of chiral systems~\cite{milovsevic2015circularly} and the tomographic reconstruction of circularly polarized high-harmonics fields~\cite{chen2016tomographic}.

In this contribution, we study the HHG in a gaseous target
driven by two single-color orthogonal polarized pulses, which are time delayed from each other.
We model the nonlinear laser-matter electron dynamics through the numerical solution
of the three-dimensional time-dependent Schr\"odinger equation (3D--TDSE) in the single-active-electron approximation. We further analyze the properties in the polarization of the generated
APT as well as its temporal anatomy by invoking signal processing and classical tools. We will show a procedure 
to twist the APT by producing a train in which each of the attosecond pulses has a well-defined linear polarization, but with a different polarization angle between them. Due to the characteristics of our synthesized polarization pulse, this technique, in the temporal domain, is similar to the bichromatic counter-rotating circularly polarized pulses, but with the main advantage of using a simpler experimental setup, i.e.~only a single-color driving pulse allows controlling the polarization angle between the attosecond pulses in the train.

\section{Results}
\subsection{Amplitude-polarized pulses}
When a linearly polarized laser pulse with an electric field $\bm{E}(t)$, carrier
frequency $\omega_0$, field envelope $f(t)$ and global phase
$\phi$, incident upon a birefringent medium
with its polarization direction to a given angle with respect to the optical axis, the field components
$E_x(t)$ and $E_y(t)$ travel with different velocities in such a medium. Thus, a temporal delay $\tau$ is generated between them. In particular, for a polarization direction at an angle of $45^{\circ}$, this field can be mathematically described as 
$\bm{E}(t)
 = E_x(t)\;\check{e}_{x}+ E_y(t)\;\check{e}_{y} =
E_0\;f(t)\;\mathrm{cos}(\omega_0t + \phi_x)\;\check{e}_{x} + E_0\;f(t -\tau )\;\mathrm{cos}(\omega_0 t + \phi_y)\;\check{e}_{y}$, where $E_0$ is the laser electric field peak amplitude, and the phases $\phi_x$ and $\phi_y$ take into account the global phase of the field $\phi$. In general,
when the phase difference $\Delta \phi = \phi_y - \phi_x = - \omega_0\tau \neq  n\pi \; (n \in \mathbf{Z})$, it is
said that the pulse has circular ($\Delta \phi= \pm\;\pi/2$) or elliptical (any other value of $\Delta \phi$) 
polarization, with ellipticity $\epsilon = 1$ ($0 < \epsilon < 1$) 
for circular (elliptical) polarization. It should be noted that in the considered case ($\Delta \phi  \neq  n\pi$), the pulse has a time-dependent ellipticity $\epsilon(t)$, since
the field envelope has a temporal delay $\tau$. One question immediately arises:
How fast is it the variation of $\epsilon(t)$? The answer can be formulated as follows: given that $\epsilon(t)$ depends on the bandwidth of the pulse $\Delta \omega$, the variation will be fast or slow depending on whether $\Delta \omega\approx\omega_0$ or $\Delta \omega\ll\omega_0$, respectively. 

On the other hand, when the phase difference is $\Delta\phi = n\pi$, the mathematical description of the field results in 
$\bm{E}(t)
= E_0\;f(t)\;\mathrm{cos}(\omega_0t + \phi)\;\check{e}_{x} + E_0\;f(t-\tau )\;\mathrm{cos}(\omega_0t + n\pi + \phi)\;\check{e}_{y} =
E_0\;(f(t)\;\check{e}_{x} + f(t-\tau)\;\check{e}_{y})\;\mathrm{cos}(\omega_0t + \phi)$, with $n$ even, and $\bm{E}(t) =
E_0\;(f(t)\;\check{e}_{x} - f(t - \tau)\;\check{e}_{y})\;\mathrm{cos}(\omega_0t + \phi)$ if $n$ is odd. It is worth mentioning .that the electric field must be written in this way because the pulse has a bandwidth $\Delta\omega\neq0$. Contrariwise, for a monochromatic field, a temporal translation $\tau = \frac{n\pi}{\omega_0}$ does not change the amplitude ratio between the two orthogonal polarizations.

Experimentally, the polarization scheme described above was presented and synthesized in Ref.~\cite{karras2015polarization},
to control the angular momentum orientation of N$_2$ molecules. Later on, other molecular scenarios were studied~\cite{tutunnikov2018selective,yuan2020generation,lin2020spatiotemporal,pan2022low,prost2018polarization,xu2020echoes}.  This polarization deserves a different classification from linear, circular or elliptical and, taking into account its mathematical expression, we will call it \textit{amplitude polarization}
(AP). However, its denomination 
is not unified in the literature, and different authors refer to it in different ways, 
namely, unidirectional polarization \cite{yuan2020generation}, twisted polarization \cite{tutunnikov2018selective}, polarization-shaped pulse \cite{lin2020spatiotemporal}, and polarization-skewed \cite{pan2022low,ji2019timing}. 

\begin{figure}[h!]
\includegraphics[width=1\textwidth]{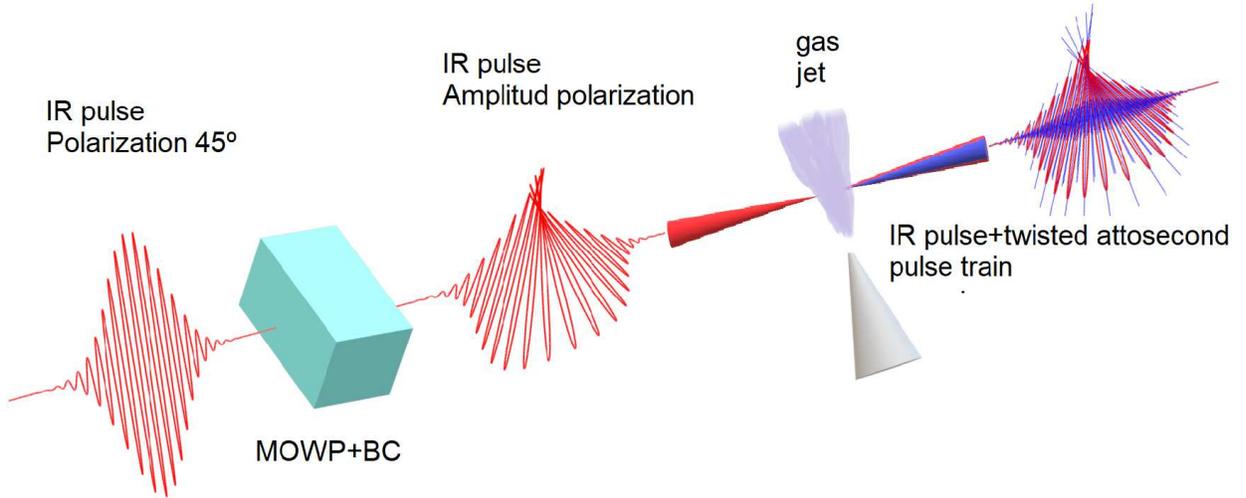}
\caption{\textbf{Schematic representation of the experimental setup} The IR--AP pulse is generated by a multiple-order wave plate (MOWP) and a Berek compensator (BC). These two optical elements are placed in the path of a single beam. The input IR pulse is linearly-polarized at $45^{\circ}$ with respect to the optical axis of the
MOWP. The synthesized IR--AP crosses a supersonic atomic gas jet, and high-order harmonics are generated, whose polarization is controlled by the properties of the IR--AP. The resulting twisted APT is then filtered and it can be used to study elementary atomic or molecular processes in its natural time scale.}\label{fig:fig1}
\end{figure}

In what follows, we will show how this AP field can be used to twist an APT. 
In Fig.~\ref{fig:fig1}, it can be seen a schematic representation of the experimental setup necessary to obtain the AP pulses and, through the HHG in a gaseous system, the twisted APT. Firstly, the IR--AP pulse is generated by a multiple-order wave plate (MOWP) and a Berek compensator (BC), 
which are placed in the path of a single beam, linearly-polarized at $45^{\circ}$ with respect to the optical axis of the
MOWP~\cite{karras2015polarization}. This configuration presents a substantial advantage over other polarization synthesis schemes for the generation or manipulation of attosecond pulses, where two-color pulses and a complex interferometric system, are required~\cite{rego2020trains,jimenez2018attosecond,fleischer2014spin,dorney2017helicity,chen2016tomographic,fan2015bright}.  In our proposal, the previously synthesized IR--AP crosses a supersonic gas jet, and high-order harmonics are generated, whose polarization is sculpted by the properties of the IR--AP. 
The resulting APT, after appropriate filtering, can then be used to drive an atomic, molecular, or solid sample, and track the multidimensional electron dynamics with attosecond time resolution.

\subsection{Twisted attosecond pulse trains}
Our aim here is to understand the underlying physics of the interaction of the IR--AP pulses with an atomic target. To this end, we will compute the HHG spectra, for different values of the relevant field parameters, on a prototypical atomic system. 

We can start by defining the AP electric field $\bm{E}(t)$ through the explicit expressions of its components $E_x(t)$ and $E_y(t)$. For instance, assuming a Gaussian field envelope $f(t)$, they are written as:
\begin{subequations}
\begin{eqnarray}
E_x(t) &=&  E_0\;e^{-2\ln(2)\left(\frac{t+\tau/2}{\Delta t}\right)^2}\;\mathrm{cos}(\omega_0(t + \tau/2 ) + \phi) \label{e1a}\\  
E_y(t) & = & E_0\;e^{-2\ln(2)\left(\frac{t - \tau/2}{\Delta t}\right)^2}\;\mathrm{cos}(\omega_0(t - \tau/2 ) + \phi), \label{e1b}
\end{eqnarray}
\end{subequations}
%
where $\Delta t$ is the temporal full width at half maximum (FWHM). For simplicity, the temporal delay $\tau$ is expressed symmetrically around $t=0$. 

Based on the expressions in Eqs.~\eqref{e1a} and ~\eqref{e1b}, our numerical calculations were performed for a value $E_0 = 0.053$ a.u., corresponding to a laser intensity $I_0 =10^{14}$ W/cm$^2$, and a wavelength $\lambda = 800$ nm ($\omega_0=0.057$ a.u.). We have analyzed the interaction between the IR--AP pulses both in the multi- and few-cycle regimes. For the first case, we considered pulses with an FWHM of $\Delta t=5$ opt. cycles (1 opt. cycle $=\frac{2\pi}{\omega_0} \approx 2.7$ fs). For the second one, we have set the FWHM to $\Delta t=2$ opt. cycles. By means of the 3D--TDSE in the single active electron (SAE) approximation (see Refs.~\cite{murakami2013high,murakami2017erratum} for details), we computed the $x$ and $y$ components of the HHG spectrum, driven by $\bm{E}(t)$. As a prototypical atomic target, we consider hydrogen (H), with an ionization potential $I_p = 0.5$ a.u. Any other atomic target can be adequately modeled by tuning the parameters of the associated model potential.  
The temporal delay $\tau$ is chosen to be $\tau=\Delta t$. For this choice, the peak strength of the synthesized AP field, $\left | \bm{E}(t) \right |_{\max}$, remains almost constant and equal to $E_0$, in the central region of the pulse (see Supplementary Information).

In Fig.~\ref{fig:fig2} we show the results obtained by the 3D-TDSE for an AP pulse with a temporal FWHM of $\Delta t=5$ opt. cycles and a global phase $\phi=0$. Figures \ref{fig:fig2}(a) and \ref{fig:fig2}(b) show the projection of the harmonic radiation in the $x$ and $y$ axis, respectively (see Supplementary Information). Meanwhile, panels (c) and (d) depict the wavelet analysis of the above-cited spectra and, superimposed, the respective electric field components $E_x(t)$ (white solid line) and $E_y(t)$ (red solid line). 

\begin{figure}[h!]
\includegraphics[width=0.9\textwidth]{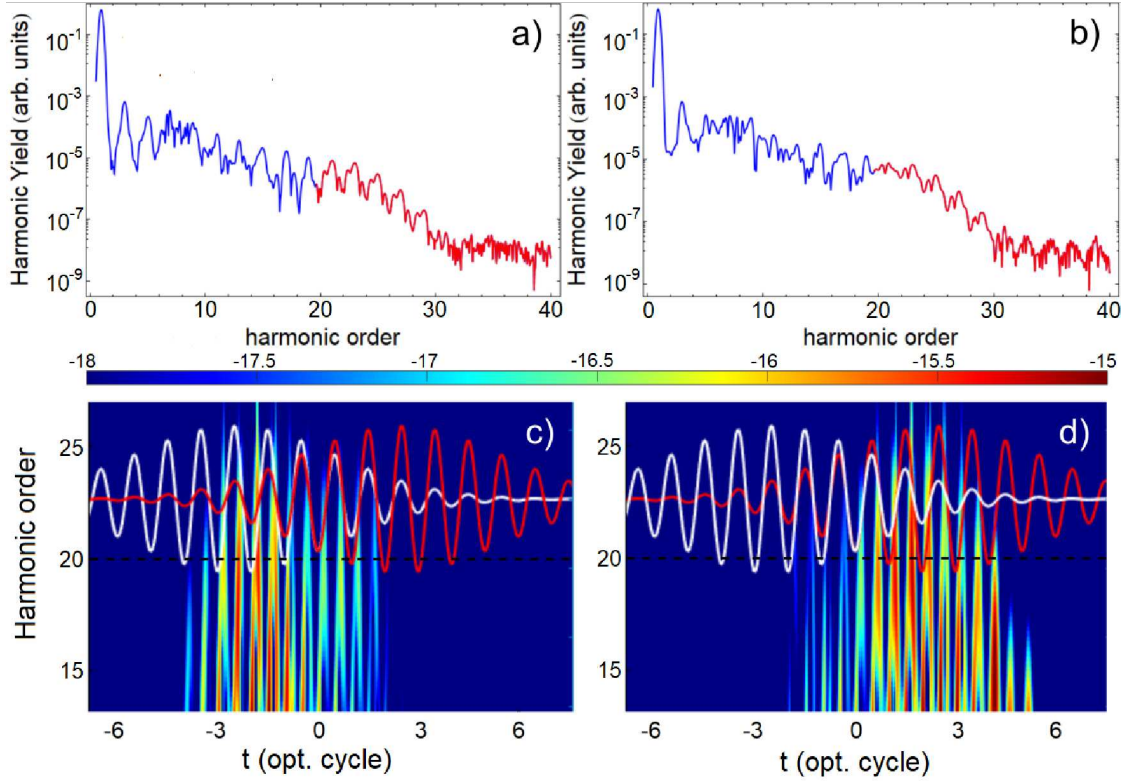}
\caption{\label{fig:fig2}\textbf{High-order harmonic generation driven by amplitude-polarization (AP) multi-cycle pulses.} Panels (a) and (b): HHG spectra generated by the field components $E_x(t)$ and $E_y(t)$, respectively. The FWHM of the AP pulse is
 $\Delta t=5$ opt. cycles and the global phase is $\phi=0$. The red regions in the spectra represent the portion selected to obtain the twisted APTs (see text for details). Panels (c) and (d): wavelet analysis obtained from the dipole accelerations $a_x(t)$ and $a_y(t)$, respectively. The $x$ ($y$) component of the AP pulse is superimposed with white (red) solid lines. The black dashed line at the 20$^{\mathrm{th}}$ harmonic defines the lower limit of the selected region.}
\end{figure}

The core of our results is shown in Fig.~3, where the temporal structure of the twisted APT can be seen. This temporal representation was obtained 
after spectrally filtering the HHG of Figs.~\ref{fig:fig2}(a) and \ref{fig:fig2}(b) above 
the 20th-harmonic. Then, only the regions of the spectrum colored in red in Figs.~\ref{fig:fig2}(a) and \ref{fig:fig2}(b) are depicted, which in turn correspond to the region above the dashed black line in Figs.~\ref{fig:fig2}(c) and \ref{fig:fig2}(d).
This spectral filtering allows us to select the so-called cut-off region,
where the spectral
phase of the harmonics is almost constant and the respective attosecond pulses can be considered Fourier-limited. On the contrary, it is well-known that in the plateau region of the HHG spectrum,
due to the different recombination times of different electron trajectories, i.e.~different emission times of the radiation (see the wavelet analysis in Figs.~\ref{fig:fig2}(c) and \ref{fig:fig2}(d)), the generation of the attosecond pulses 
has an intrinsically spectral phase, the so-called atto-chirp. In general, this phase can be experimentally compensated by a metallic foil~\cite{chini2014generation}.

\begin{figure}[ht!]
\includegraphics[width=.6\textwidth]{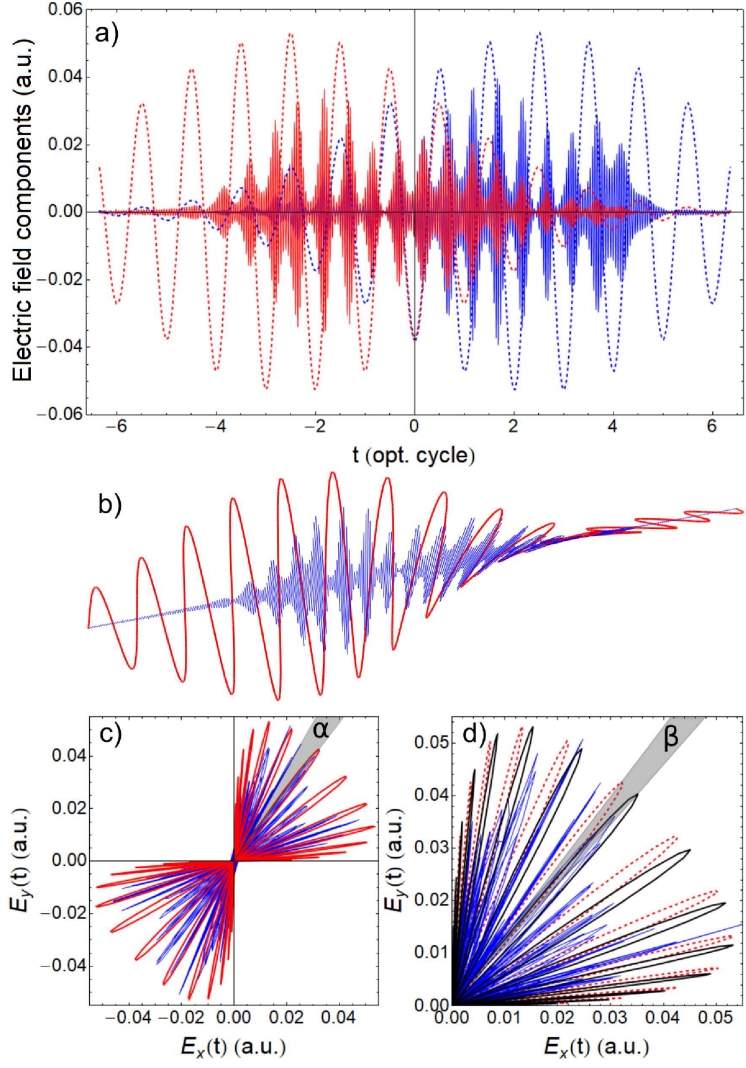}
\caption{\label{fig:fig3}\textbf{
Temporal anatomy of the twisted APT: multi-cycle pulses regime.} 
(a) Projection of the twisted APT in the $x$--$y$ plane (solid line) and the field components $E_x(t)$ and $E_y(t)$ of the AP pulse (red and blue dashed lines, respectively). The results that are shown correspond to those of Fig.~\ref{fig:fig2}, in the selected region of the spectra. (b) A 3D--representation of the twisted APT (blue solid line) and the laser electric field of the AP pulse (red solid line). (c) Lissajous figures of the laser electric field of the AP pulse (red solid line) and the twisted APT (blue solid line). The shadow region depicts the angular spacing $\alpha$,
between the polarization direction of two adjacent attosecond pulses. (d) Same as panel (c) but for a driving AP field with a global phase $\phi=\pi/2$ (only the first quadrant is plotted). The laser electric field of the AP pulse with a global phase $\phi=0$ is also shown (red dashed line). The shadow region depicts the angular spacing $\beta$, between the polarization direction of the two AP pulses, with global phases $\phi=0$ and $\phi=\pi/2$.}
\end{figure}

Let us now aim to dissect the twisted temporal anatomy of the APT. For this purpose we have plotted, in Fig.~\ref{fig:fig3}(a), the components of the field $\bm{E}(t)$ (dashed lines) and the twisted APT (solid lines). The $x$-components ($y$-components) are shown in red (blue) color. From there, 
it can be seen that the attosecond pulses are generated every half cycle of the AP pulse and, what is most important, the relative amplitude between the $x$ and $y$ components of the attosecond pulses in the train, changes from one pulse to the next. This indicates that the polarization 
between two adjacent attosecond pulses is different. Furthermore, the full temporal 3D--representation of the twisted APT can be seen in Fig.~\ref{fig:fig3}(b) (blue solid line), where we have also included the laser electric field of the AP pulse (red solid line). From this figure, the temporal evolution of each attosecond pulse polarization in the train and how it rotates is clearly observed by following the temporal evolution of the AP pulse.

For a better visualization of the polarization features of  the individual attosecond pulses, in Fig.~\ref{fig:fig3}(c) we show the Lissajous curves of both, the laser electric field of the AP pulse (red solid line) and the electric field of the twisted APT (blue solid line). Two key observations
can be extracted from this figure: First, each attosecond pulse has almost perfect linear polarization. Second, the polarization direction of each attosecond pulse follows the orientation of the ``petals" that describe the polarization of the AP electric field. 
It is possible to obtain the difference in the polarization direction between two adjacent attosecond pulses, $\alpha$ (see the shadowed region), which in this case is $\approx 7.5^\circ$.    


Finally, from Fig.~\ref{fig:fig3}(d), we can analyze how the previous results vary by changing the global phase of the driving AP pulse by $\pi/2$. 
Considering that the polarization direction of the attosecond pulses is given by the position of the petals, for a clear comparison, we show the Lissajous curves of the electric fields only in the first quadrant. Here, an AP pulse with global phase $\phi=0$ ($\phi=\pi/2$) is represented by a dashed red (solid black) line, while the twisted APT is depicted by a blue solid line. The change in the polarization direction 
between these AP fields, with different global phases, is indicated as $\beta$ (see the shadowed region). In this case $\beta\approx 3.9^\circ$, i.e~$\beta\approx\alpha/2$. Thus, when the global phase $\phi$ changes in $\pi/2$, the APT rotates, globally, at an angle equal to $\beta$.

\begin{figure}[h!]
\includegraphics[width=0.6\textwidth]{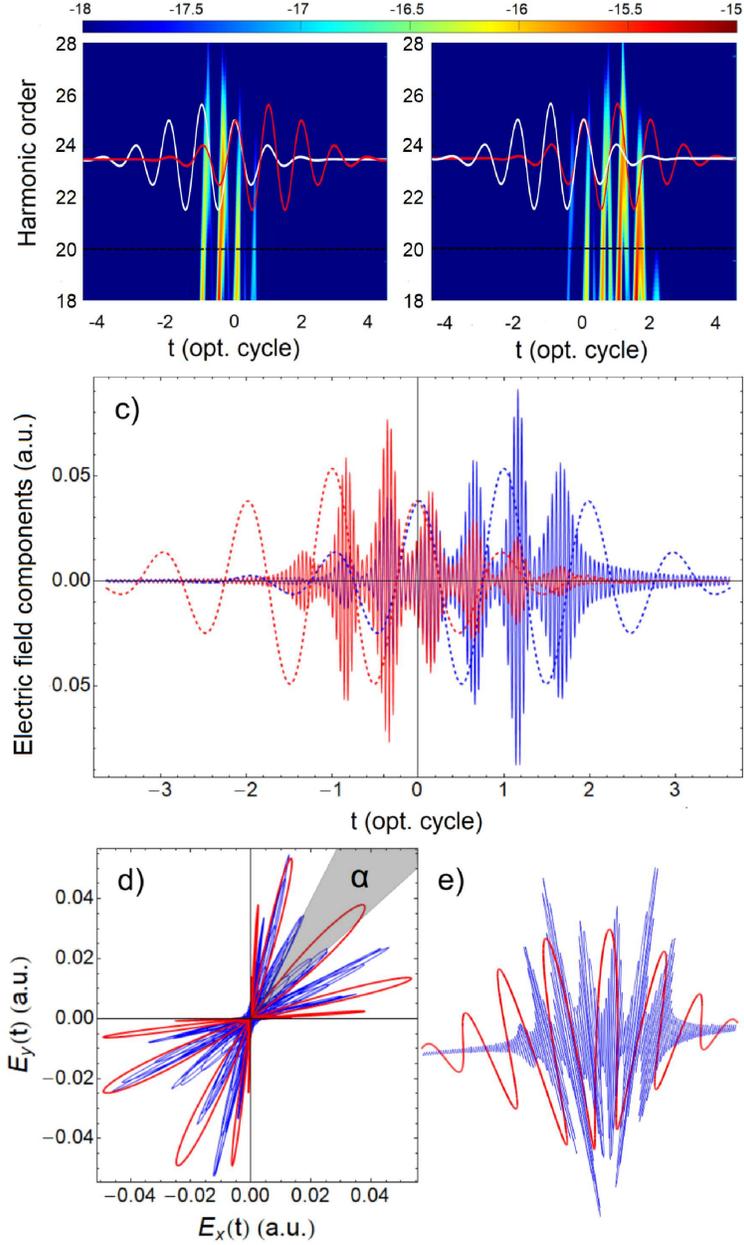}
\caption{\label{fig:fig4} \textbf{
Temporal anatomy of the twisted APT: few-cycle pulses version.} Same that Figs. \ref{fig:fig2}(c) and \ref{fig:fig2}(d) (in this figure, panels (a) and (b), respectively), and Figs. \ref{fig:fig3}(a), \ref{fig:fig3}(b) and \ref{fig:fig3}(c) (in this figure, panels (c), (d) and (e), respectively), but obtained from an AP pulse with a temporal FWHM of $\Delta t=2$ opt. cycles.}
\end{figure}

All the analysis presented above was made in the multi-cycle regime. In order to complete the picture, in the following we examine the dynamics for a few-cycle pulse. In Fig.~\ref{fig:fig4} we show our results for an AP pulse with an FWHM of $\Delta t=2$ opt. cycles. For this FWHM, the angular spacing between two adjacent petals in the AP pulse is significantly larger than in the previous multi-cycle case. In order to make a one-to-one comparison, in Figs.~\ref{fig:fig4}(a) and \ref{fig:fig4}(b) we show the wavelet analysis coming from the associated HHG spectra in this few-cycle instance. The $x$ and $y$ components of the AP electric field $\bm{E}(t)$ are superimposed
in white and red solid lines, respectively. Likewise, in Fig.~\ref{fig:fig4}(c) we plot both the electric field components of the AP field  (dashed line) and the APT (solid line). Here, the $x$-components ($y$-components) are shown in red (blue) color. As in the previous case, the relative amplitude between the components of the attosecond pulses changes from one pulse to the next. From the Lissajous curve (Fig.~\ref{fig:fig4}(d)) a slightly less ellipticity of the attosecond pulses is observed in relation to that observed in Fig.~\ref{fig:fig3}(d), and a larger angle $\alpha$ between the polarization direction of two adjacent pulses ($\alpha\approx 20^\circ$). 
The full temporal evolution of the twisted APT is then shown in Fig.~\ref{fig:fig4}(e). As in the multi-cycle regime, the HHG spectra were spectrally filtered, and we depicted here only from the 20th-harmonic onwards (see the dashed black line in Figs.~4(a) and 4(b)). 

 Further analysis of the value of $\alpha$ as a function of the temporal FWHM and the time delay $\tau$, can be seen in SM. As a summary, we can mention that $\alpha$ increases almost linearly whit $\tau$, which gives us a clear idea of how to manipulate the change in the polarization direction between two adjacent attosecond pulses from the synthesis of the driving IR field.
Furthermore, the calculation of the recollision angles of
the ionized electrons~\cite{murakami2013high}, as well as a classical analysis of the trajectories of the electrons in the continuum, are included in the SM.

\section{Discussion}

In this work, we have introduced a straightforward procedure to generate an APT in which each of the attosecond pulses has a well-defined linear polarization, but different between them.
This appealing feature of the APT is achieved through the manipulation of the polarization state of the driving IR field. Here, by introducing a variable time delay between two identical copies of an IR pulse, we obtain an AP pulse.  Classically speaking, the generation of this twisted APT is rooted in the small deviation of the trajectories of the electrons in the continuum by the AP pulse, due to the temporal change in its polarization state. This makes the electrons recombine at different angles, where the ``short" trajectories slightly deviate in relation to the ``long" ones. Furthermore, this deviation is more pronounced with few-cycle driving pulses, which allows for reaching larger rotation angles. Thus, this small deviation of the electrons' trajectories in relation to the linear polarization pulse, allows us to extend the current results by employing different conventional sources to obtain an APT, such as high-energy laser sources, at a high repetition rate, or with different wavelengths. Moreover, our scheme is quite versatile and robust, as it can be implemented in both the multi- and few-cycle regimes.

The concrete possibility to manipulate the polarization angle between the attosecond pulses in an APT would bring attosecond-based spectroscopy techniques to another, more advanced, level~\cite{jiang2022atomic,gong2022asymmetric,busto2019fano,beaulieu2017attosecond,joseph2020angle}. For instance, it would be possible to generate photoelectrons within a well-defined angular range or design complex pump-probe techniques. Just to cite a few examples, we could think of several pump-probe schemes, namely an IR pulse/a twisted APT, an AP pulse/a twisted APT, or a twisted APT/twisted APT. The latter would be an XUV pump-XUV probe scheme, but when each of the pulses is a train of twisted isolated attosecond pulses. In addition, because we can manipulate the polarization angle between two adjacent pulses in the APT, one can imagine sophisticated attosecond-based spectroscopy techniques using a single beam. Beyond photoelectron spectroscopy, this twisted APT can be useful to (i) study or drive highly anisotropic systems, for instance, systems where there are preferential directions, such as molecular systems, low-dimensional crystalline structures, etc.~\cite{li2020attosecond} and (ii) characterize multidimensional laser fields in the time domain utilizing the well-known streaking technique, mainly used for linearly polarized fields~\cite{streaking2004}, amongst other.  

Likewise, our approach presents clear advantages compared to other isolated attosecond pulse generation techniques (polarization gating, lighthouse, amplitude gating, or ionization gating~\cite{chini2014generation}). First, the possibility to work with multi-cycle pulses (in most of the isolated attosecond pulses generation techniques few/single-cycle pulses are needed). Second, the synthesis of the AP pulses is experimentally straightforward: only a  single color pulse without the need for an involved interferometric system is necessary, i.e, we work in a single beam path. Third, if the spectral phase of the harmonics is compensated, for example with a metallic foil, it would be possible to use a large spectral bandwidth, generating ultrashort attosecond pulses.    

Finally, let us emphasize the instrumental role of the global phase $\phi$ as an additional `knob' in our proposed scheme. For instance, we have proved that a change of the global phase in $\pi/2$ of the driving AP pulse echoed by a global rotation of the twisted APT. This implies that it is necessary to stabilize the global phase (carrier-envelope phase) of the AP pulse for pump-probe experiments or if we aim for coherent control in some processes. On the contrary, this global rotation of the twisted APT can be useful to detect the change in the global phase for the AP pulse, both in the few-cycle and multi-cycle regimes.


\section{Methods}

To find the quantum dipole acceleration $\bm{a}(t)
 = a_x(t)\;\check{e}_{x}+a_y(t)\;\check{e}_{y}$, we have numerically solved the 3D--TDSE~\cite{murakami2013high,murakami2017erratum}, in the dipole approximation, by expanding the active electron wavefunction in spherical harmonics $Y^{m}_l(\theta,\phi)$:

 \begin{equation}
\Psi(\bm{r},t)=\sum \frac{R^{m}_l(\bm{r},t)}{r}Y^{m}_l(\theta,\phi)\label{eM1},
\end{equation}
where $\theta$ and $\phi$ are the polar and azimuthal angles. Through the Ehrenfest’s theorem, the $x$ and $y$ components of the acceleration operator $\hat{\bm{a}}(t)$ can be found as

\begin{subequations}
\begin{eqnarray}
\hat{a}_x & = &  -[\hat{H},[\hat{H},\hat{x}]]\;=\;\left[\hat{H},\frac{\partial}{\partial r}\right]\;\mathrm{cos(\theta) },\label{eM2a}\\ %
\hat{a}_y & = & -[\hat{H},[\hat{H},\hat{y}]] \; = \;\left[\hat{H},\frac{\partial}{\partial r}\right]\;\mathrm{sin(\theta)cos(\phi)}, \label{eM2b}
\end{eqnarray}
\end{subequations}
where $\hat{H}$ is the full Hamiltonian. Thus, the components of $\bm{a}(t)$ can be obtained as: 
$a_x(t)=\expval{\hat{a}_x}{\Psi(\bm{r},t)}$ and $a_y(t)=\expval{\hat{a}_y}{\Psi(\bm{r},t)}$.
The projections in the $x-y$ plane of the harmonics spectra amplitudes $|\tilde{a}_x(\omega)|$ and $|\tilde{a}_y(\omega)|$ (Figs.~2(a) and 2(b)), and the respective spectral phases, $\gamma_x(\omega)$ and $\gamma_y(\omega)$, are then calculated by the Fourier transforms

\begin{subequations}
\begin{eqnarray}
\tilde{a}_x(\omega) &=&|\tilde{a}_x(\omega)|e^{i\gamma_x(\omega) }  \;=\;\frac{1}{(t_f-t_i)\;\omega^2}\int_{t_i}^{t_f}e^{-i\omega t}a_x(t) \;dt \label{eM3a}\\  
\tilde{a}_y(\omega) &=&|\tilde{a}_y(\omega)|e^{i\gamma_y(\omega) } \; = \;\frac{1}{(t_f-t_i)\;\omega^2}\int_{t_i}^{t_f}e^{-i\omega t}a_y(t) \;dt,  \label{eM3b}
\end{eqnarray}
\end{subequations}
where the times $t_i$ and $t_f$ define the integration window, both to solve the 3D--TDSE and to evaluate these integrals. For the first case, of the two analyzed in this work, where the FWHM of the pulse is $\Delta t=5$ opt. cycles, the numerical calculations were performed for $t_i=-2757$ a.u. and $t_i=2757$ a.u., with a time step of 1 a.u. In the second case, where the FWHM of the pulse is $\Delta t=2$ opt. cycles, the integration times are $t_i=-1103$ a.u. and $t_i=1103$ a.u., with a time step of $1$ a.u.   

Once we obtain $\bm{a}(t)$, a wavelet analysis can be performed by computing the Gabor transform, for both the $x$ and $y$ components, as follows:

\begin{equation}
G_{x,y}(\omega,t)=\left |\int  a_{x,y}(t')\;\frac{\mathrm{exp}[-(t-t')^2/2\sigma^2]}{\sigma\sqrt{2\pi}}\;\mathrm{exp}(i\omega t')\;dt'\right |^2.\label{eM4}
\end{equation}
A value of $\sigma=1/(3\omega_0)$ allows achieving an adequate balance between the time and frequency resolutions. $G_{x,y}(\omega,t)$ then gives us access to study the time dynamics behind the HHG process and compare it with classical simulations.

The $x$--$y$ components of the electric field of the different APT, $E_{XUV,x}(t)$ and $E_{XUV,y}(t)$, are calculated by filtering the harmonic spectrum, and starting from the 20$^{\mathrm{th}}$ ($\omega_i=20$) harmonic order:

\begin{subequations}
\begin{eqnarray}
E_{XUV,x}(t) &=&\int_{\omega_i=20\omega_0}^{\infty}e^{i\omega t}\tilde{a}_x(\omega) \;d\omega, \label{eM5a}\\  
E_{XUV,y}(t) &=&\int_{\omega_i=20\omega_0}^{\infty}e^{i\omega t}\tilde{a}_x(\omega) \;d\omega.  \label{eM5b}
\end{eqnarray}
\end{subequations}

\section*{Acknowledgements}

M. F. C.~acknowledges financial support from the Guangdong Province Science and Technology Major Project (Future functional materials under extreme conditions - 2021B0301030005). E. G. N.~acknowledges to Consejo Nacional de Investigaciones Cient\'ificas y T\'ecnicas (CONICET). F. A. V.~acknowledges to Comisi\'on de Investigaciones Cient\'ificas de la Pcia. de Buenos Aires.   




\bibliography{main}

\end{document}